\documentclass[12pt]{article}
\usepackage{hyperref, graphicx, color, amssymb, amsmath, cite}
\usepackage[margin=1.0in]{geometry}
\usepackage{cmap}
\newcounter{si}

\begin{document}
\title{Chemical Characterization of Aerosol Particles Using On-chip Photonic Cavity Enhanced Spectroscopy}
\author{
        Robin Singh, Brian W. Anthony \\
        Department of Mechanical Engineering\\ 
        Institute for Medical Engineering and Science\\
        MIT.nano\\ 
        Massachusetts Institute of Technology\\
        Massachusetts Ave, Cambridge MA 02139, USA
            \and
        Danhao Ma, Lionel Kimerling, Anu Agarwal\\
        Department of Materials Science \& Engineering\\
        Microphotonics Center
        Materials Research Laboratory\\
        Massachusetts Institute of Technology\\
        Massachusetts Ave, Cambridge MA 02139, USA
}
\date{\today}
\maketitle
\begin{abstract}
We demonstrate the chemical characterization of aerosol particles with on-chip spectroscopy using a photonic cavity enhanced silicon nitride (Si3N4) racetrack resonator-based sensor. The sensor operates over a broad and continuous wavelength range, showing cavity enhanced sensitivity at specific resonant wavelengths. Analysis of the relative change in the quality factor of the cavity resonances successfully yields the absorption spectrum of the aerosol particles deposited on the resonators. Detection of N-methyl aniline based aerosol detection in the Near InfraRed (NIR) range of 1500 nm to 1600 nm is demonstrated. Our aerosol sensor spectral data compares favorably  with that from a commercial spectrometer, indicating good accuracy. The small size of the device is advantageous in remote, environmental, medical and body-wearable sensing applications. \end{abstract}

\section{Introduction}
Aerosols consist of solid or liquid particles suspended in a gas medium. They play a significant role in the physio-chemistry of the environment, climate forcing, air quality, and the transport mechanism of accidental or industrial release of biological and chemical weapons \cite{Rubin2010}\cite{Cao2017}\cite{Ibrahim2015} They play an essential role in pharmaceutical research, particularly for pulmonary drug delivery applications. For instance, many pulmonary applications use nebulizers that generate aerosols to deliver drugs within lung airways \cite{Tiwari2012}. This requires tight control on aerosol properties such as particle size and count to control the fate of aerosol deposition location in the respiratory tract. Similarly, many transdermal applications use nanoparticle-based aerosols for drug delivery \cite{Tiwari2012}. Many applications will benefit from a miniaturize and low cost techniques to characterize the microphysical properties of an aerosol.\\

Both physical and chemical characterization of aerosols are beneficial. Physical refers to estimating the particle size and count in the test medium, and the latter relates to the chemistry of the particles. For decades, the physical characterization of aerosol particles has been performed using free space optical and electrical methods that involve evaluation of light scattering and electrical mobility properties of the medium \cite{Kerker1997}\cite{Profile2015}. For chemical characterization, commonly used methods are based on Fourier Transform InfraRed (FTIR) spectroscopy \cite{Kira2016}, Raman spectroscopy \cite{Aggarwal2014}, and fluorescent imaging \cite{Jung2012}. These techniques for chemical characterization is that they are either too bulky for field testing \cite{Saito2018} or have poor sensitivity when miniaturized to a hand-held form factor \cite{Zhang2017}.\\

There is a need for a miniaturized technique to chemically characterize the aerosol particles without compromising the sensitivity of the device, while also offering the advantages of low-cost and real-time monitoring \cite{Armani2003,Wade2016,Hu2008}. Commercial portable sensors today can measure physical properties of aerosols, such as their particle size distribution and count \cite{Crilley2018}, and portable mid-IR spectrometers based on micro electro mechanical systems (MEMS) technologies can chemically characterize them, but would benefit from improved detection limits \cite{Nitkowski2008,Ozdemir2014}\cite{Castell2017}. \\

The present work develops an ultra-sensitive on-chip photonic spectroscopy platform to obtain chemical attributes of the aerosol particles. The proposed method uses a Si3N4 based micro ring resonator cavity as the sensor, to enhance light-particle interaction for improved detection limit \cite{Zhu2010,Wang2016,Tao2017,Soria2011}. The wavelength range chosen in the study is dependent on the chemical composition of the aerosol particle to be detected, because every chemical absorbs identifiably distinct bands of the infrared spectrum, called its chemical fingerprint. The current study chooses the near IR regime to detect N-methyl aniline aerosols. Such an approach can be used to develop a highly sensitive on-chip photonic aerosol spectrometer \cite{Wang2016}. \\

 \begin{figure}[h!]
	\centering
	\includegraphics[width=1\textwidth]{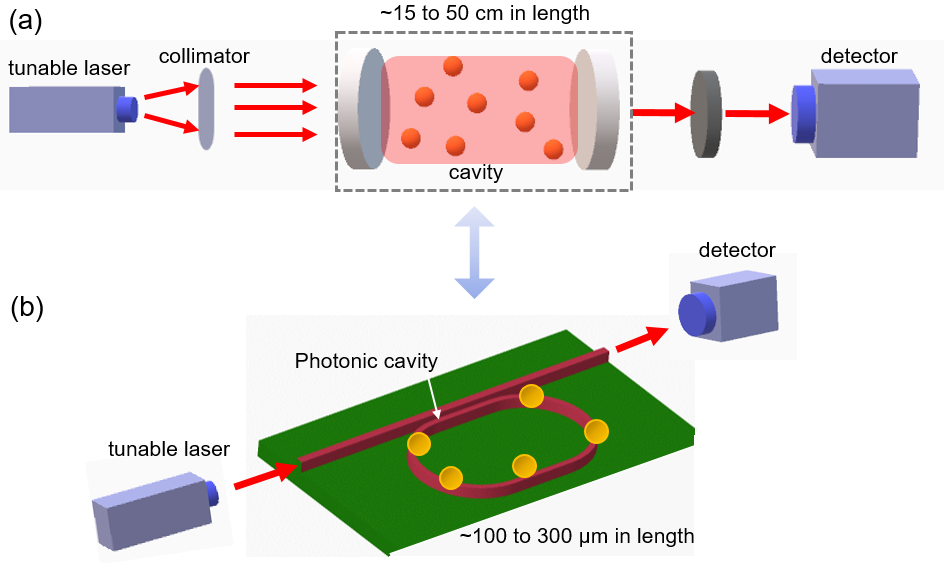}
	\caption{\textit{(a) Depiction of conventional instrumentation set up for cavity enhanced aerosol spectroscopy. The method uses a tunable light source, collimator, a cavity to allow particle-light interaction, lens, and detector.  The cavity increases the optical path of the light from the source to the detector. The cavity comprises two reflecting surfaces that allow light to bounce back and forth while interacting with the aerosol particles dispersed in the cavity (b) Depiction of on-chip photonic cavity for aerosol spectroscopy. The cavity consists of optical waveguides forming a resonator coupled with the linear (bus) waveguide. As the light couples into the bus waveguide from the tunable source, it circulates in the resonator loop multiple times. The circulation in the resonators depends on the quality factor of the cavity and enhances light-aerosol interaction. The transmittance spectrum of the cavity is obtained from the output of the detector}}
\end{figure}

\section{Photonic Device Design and Fabrication}
\subsection{Photonic Device Design }
Our approach for cavity enhanced on-chip spectroscopy of aerosol particles is based on a micro racetrack resonator with waveguides designed as racetrack loops, as shown in Fig. 1. We couple in the light from the tunable light source to a bus waveguide through edge coupling \cite{Zhu2010}. Light in the bus waveguide interacts with the photonic cavity through the gap between bus waveguide and cavity loop. When the optical path length of the cavity waveguide is an integral multiple of the input light wavelength, the light circulates within the loop multiple times and enhances the light particle interaction during spectroscopy. The coupling efficiency is determined by the gap $g$ between the bus waveguide and resonator waveguide and the coupling length $L_c$ \cite{Su2017}\cite{Wade2016}. The annotated schematic of the racetrack resonator is shown in Fig. 5c. Compared to a micro ring structure, a racetrack structure enhances the coupling efficiency due to its larger coupling length. Since free spectral range (FSR) depends inversely on the coupling length of the resonant cavity, a racetrack structure decreases the FSR \cite{Su2017}\cite{Wade2016}. Smaller FSR of the design results in greater number of input light wavelengths resonating within the cavity over a given wavelength swept range, improving the spectral resolution of the sensor.\\

The resonant cavity is designed to support wavelengths between 1500 nm to 1600 nm (NIR). To design, optimize and evaluate the photonic cavity, we simulate the microstructures using Lumerical MODE simulation software [Lumerical Incorporated, Canada]. Fig. 3 shows mode simulations and effective refractive index analysis that is used to optimize the width of the waveguide within the cavity for supporting a single mode. The simulation model consists of the silicon substrate with a buried oxide (BOX) which is a 3 $\mu$m thick silicon dioxide layer. On top of the BOX layer are the silicon nitride (Si3N4) based waveguide structures of the sensor. Given a single-mode Si3N4 thickness of 400 nm, we calculate the waveguide width that continues to support this single mode. Effective refractive index analysis is done for different widths of the waveguides (Fig. 3e). Based on simulations, the optimized cross section of a single mode waveguide is 400 nm in thickness and 800 nm in width for 1.5 to 1.6 $\mu$m wavelength. The first two (TE and TM) modes supported by the cavity waveguide are presented in Figs. 3c and 3d. The aerosol particles interact with the evanescent wave tail in the vicinity of the cavity waveguide.

\subsection {Fabrication}
We fabricate the device with the process flow as shown in Fig. 2. A low pressure chemical vapor deposition (LPCVD) system is used to deposit 400 nm thick silicon nitride (Si3N4) layer on 6-inch silicon dioxide wafer (3-micron oxide on Si substrate). These thermal oxide wafers are procured from Wafer Pro LLC, CA.  Micro-racetrack resonators and waveguides are designed and patterned on silicon nitride-on-insulator substrate via photolithography and followed by reactive ion etching to define the geometry of the on-chip sensing components. Fluorine chemistry with a gas mixture of CHF3 and CF4 is used in the dry etching step.  Fig. 3 shows the SEM image of the fabricated resonators with its dimensions. 
\begin{figure}[h!]
	\centering
	\includegraphics[width=1\textwidth]{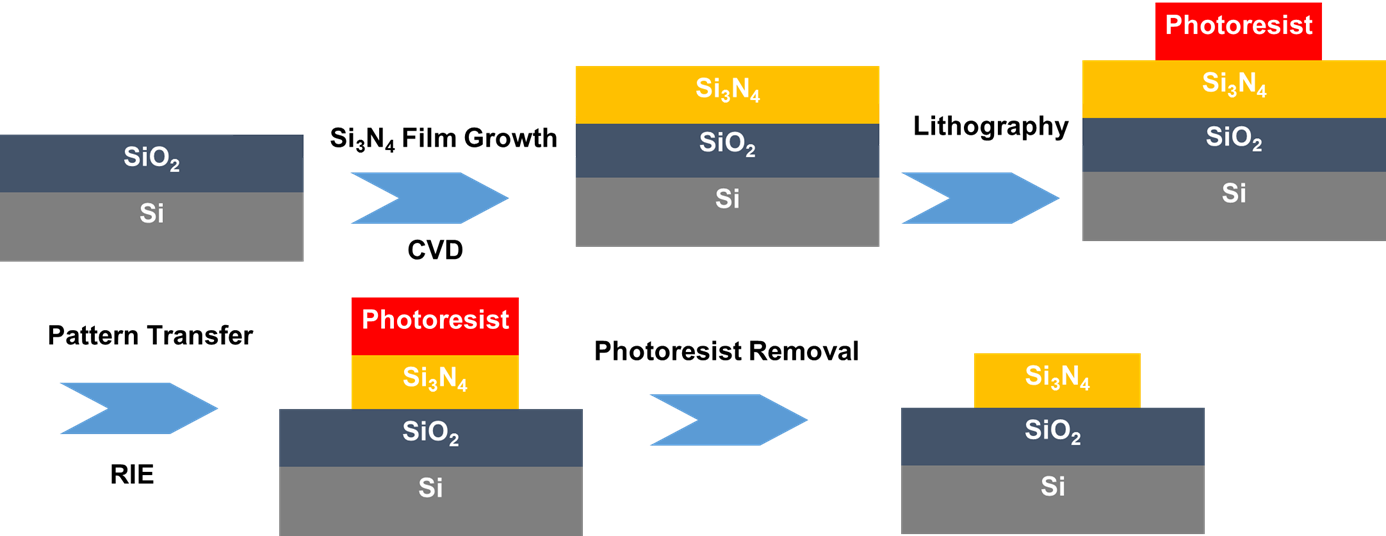}
	\caption{\textit{Schematic showing the process flow to fabricate the Si3N4 based micro race track resonators used in the study. We deposit 400 nm thick Si3N4 on SiO2+Si wafer. We use photolithography to pattern the waveguide structures, followed by pattern transfer using reactive ion etching process. Once the device is patterned, we remove the photoresist by cleaning the wafer with plasma treatment.}}
\end{figure}

Fig. 4 shows the schematic of the setup to detect aerosol particles. N-methyl aniline based aerosols are generated with a TSI 3076 atomizer [TSI Incorporated, Shoreview, USA] which uses a compressed air-based atomization technique to produce liquid aerosol particles, which are then passed through a TSI diffusion dryer [TSI Incorporated, Shoreview, USA] to dry the aerosol mixture. We use a handheld pump to maintain a constant flow rate of aerosol particles within the sensor channel. In addition, a differential mobility analyzer (TSI-3080) can be connected to the diffusion dryer to select a specific particle size at the output. 

\begin{figure}[h!]
	\centering
	\includegraphics[width=1\textwidth]{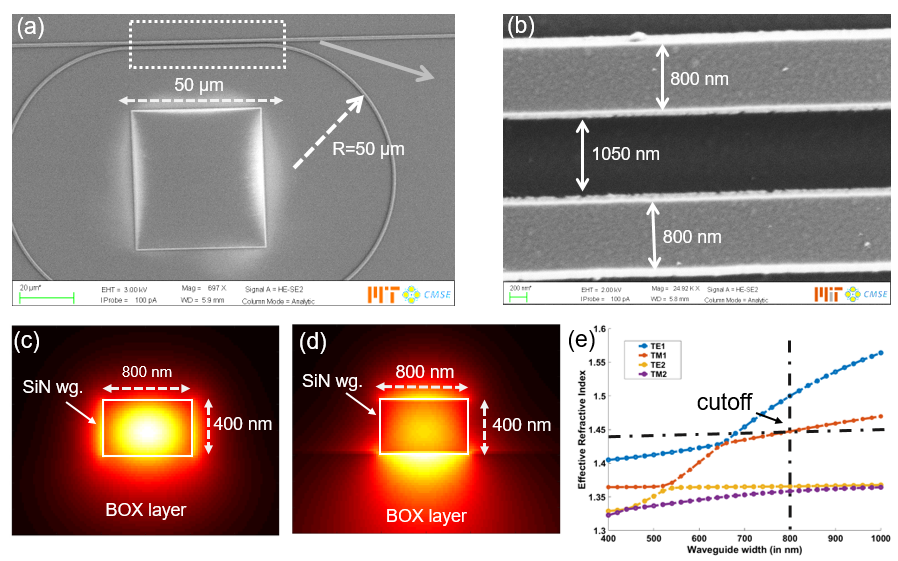}
	\caption{\textit{Design and characterization of the sensor. (a) SEM image of the on-chip photonic cavity in the aerosol sensor. (b) Expanded SEM view of the race track resonator showing critical coupling dimensions of racetrack and bus waveguides in the aerosol sensor. (c, d) MODE analysis of the resonator waveguide, with c and d showing the TE and TM modes respectively. (e) Waveguide width design to obtain a single mode resonator, by performing a waveguide width sweep given a constant thickness of 400 nm.}}
\end{figure}

\begin{figure}[h!]
	\centering
	\includegraphics[width=1\textwidth]{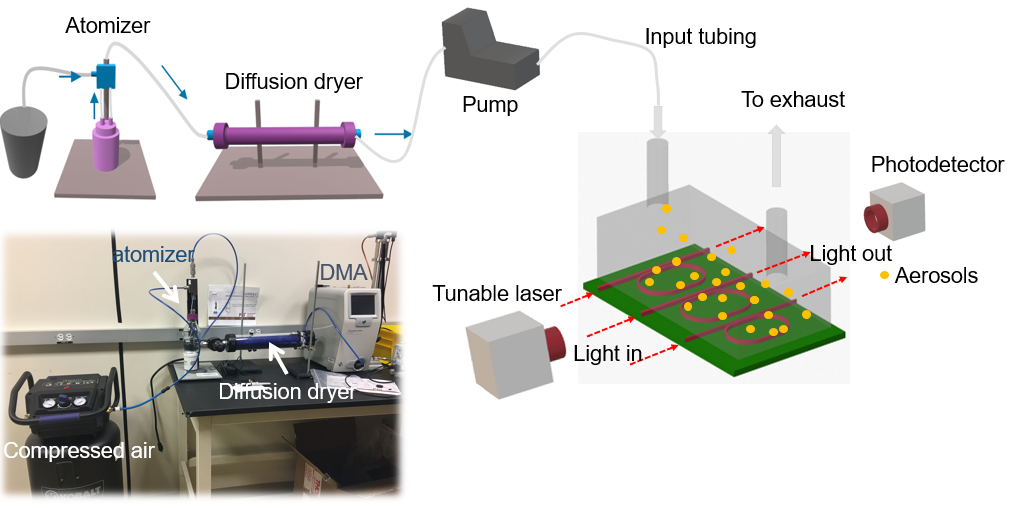}
	\caption{\textit{Schematic of the setup used in on-chip aerosol spectroscopy. It comprises of three different subsystems: (a) aerosol delivery and transport system (b) on-chip photonic sensing system (c) output pattern and spectrum acquisition system. Aerosols from the test environment (here, a constant output atomizer) are channeled into the aerosol transport system comprising a constant pressure pump, which regulates flow rate in the delivery tube.}}
\end{figure}

Our photonic sensor is edge coupled to the tunable laser and detector from Luna technologies [PHOENIX 1200, Luna Innovations Incorporated, VA, USA]. The laser has tuning capability from 1520 nm to 1600 nm wavelength. Fig. 5a shows the sensor coupled with the system. Fig. 5d to 5g show the microscopic images of the sensors with and without the aerosols. The cavity highlighted with the red arrow in Fig. 5f to 5h is used as the sensor module. 

\begin{figure}[h!]
	\centering
	\includegraphics[width=1\textwidth]{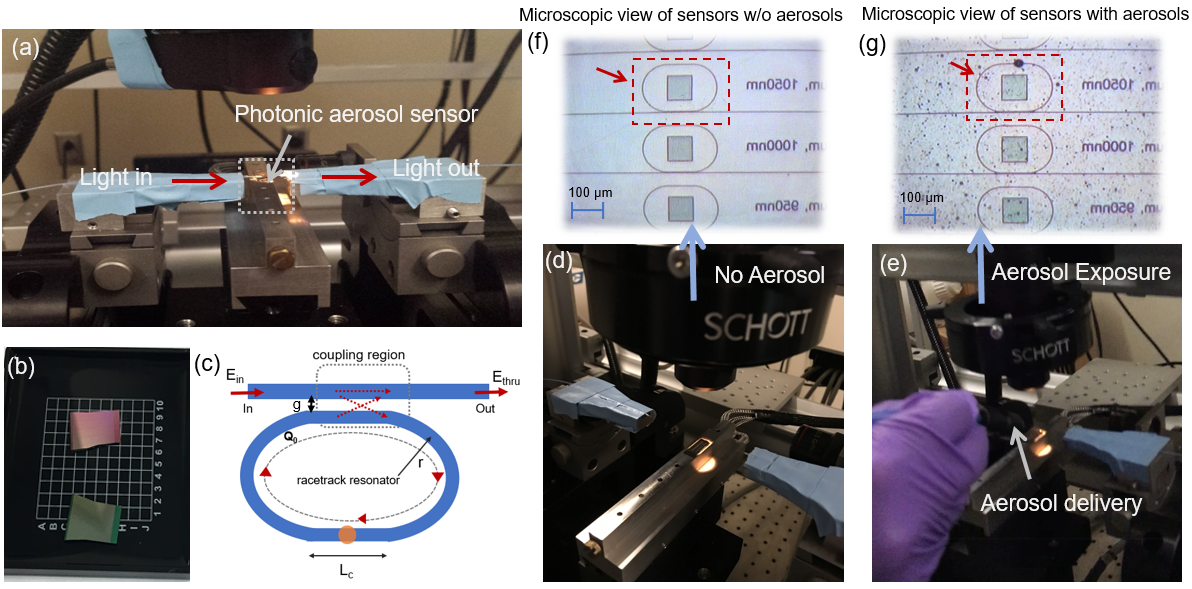}
	\caption{\textit{Photonic Aerosol Sensor. (a) light is edge-coupled in and out of the sensor device. (b) A picture of the fabricated aerosol sensor (c) Schematic of the racetrack resonator. Microscopic views of the aerosol sensor (d, e) without aerosols and (f, g) with aerosols.l}}
\end{figure}

The absorption spectrum is determined from the transmittance spectrum measured by the resonant cavity. We present a theoretical model for the racetrack cavity resonance. Fig. 5c depicts different parameters of the racetrack cavity used in the mathematical model. The round-trip length, ${L_{total}}$ , in the racetrack cavity, is given as derived in \cite{Wade2016}, 

\begin{equation}
{L_{total}} = 2{L_c} + 2\pi r
\end{equation}

Where $r$ and ${L_c}$ are the bend radius and the coupling length respectively. When the laser input is tuned over the given spectral range of interest, transmittance response of the all-pass racetrack resonant cavity is given by \cite{Wade2016}, 

\begin{equation}
{{{E_{thru}}} \over {{E_{in}}}} = {{ - \sqrt A  + t{e^{ - i\phi }}} \over { - \sqrt A {t^*} + {e^{ - i\phi }}}}
\end{equation}

where t is the straight-through coupling coefficient of the optical field, and t* is the complex conjugate of ‘t’. Here, ‘t’ is mathematically defined as 

\begin{equation}
t = \cos \left( {{{\pi L_{total}} \over {2{L_c}}}} \right){e^{i\beta L_{total}}}
\end{equation}

Where ${L_c}$ is the coupling length of the racetrack resonator, $\phi $ and A are the round-trip phase and the power attenuation, respectively. $\phi $  is defined as $\phi  = \beta {L_{total}}$, $\beta$ being propagation constant and A is the power attenuation given by ${e^{ - {\alpha _{total}}{L_{total}}}}$. The power attenuation during the propagation is given by ${\alpha _{total}} = {\alpha _{aerosol}} + \Gamma {\alpha _{wg}}$  where, ${\alpha _{aerosol}}$ and ${\alpha _{wg}}$ are absorption coefficients of the aerosol and of the Si3N4 waveguides and $\Gamma $ is the confinement factor of the waveguide that determines the fraction of the guided tail (evanescent waves) that is available outside the waveguides to interact with the aerosol particles.

One of the useful parameters that relates the absorption coefficient of the waveguides with the nature of the resonant peak is the quality factor, ‘Q’, of the resonant peak. Q is defined as \cite{Wade2016}
\begin{equation}
Q = {{2\pi {n_r}} \over {{\alpha _{total}}{\lambda _m}}}
\end{equation}
From the above equation, it can be concluded that $Q \propto {1 \over {{\alpha _{total}}}}$. Given that ${\alpha _T} = {\alpha _{aerosol}} + \Gamma {\alpha _{wg}}$, $\Gamma {\alpha _{wg}}$  remains constant for the sensor design, $Q \propto {1 \over {{\alpha _{aerosol}}}}$ . Hence, analyzing the quality factor of the resonance peaks can be used to retrieve the relative absorption of the aerosol particles at the resonant wavelengths. Here, we calculate the relative change in the quality factor (Q1-Q2)/Q2 for all resonant peaks, where Q1 and Q2 are the quality factor before and after aerosol exposure respectively.

\section{Results and Discussions}

Fig. 6e shows the transmission spectrum of the resonant cavity with an exposure time of 20, 40, 80 and 120 seconds to the N-methyl aniline based aerosols. Once the aerosol distribution on the sensor reaches a steady state in the flow rate, we measure the transmission spectrum of the cavity over the wavelength range of 1520 nm-1600 nm. The micrographs at four different exposure times are shown in Fig. 6a-d. Fig. 6e shows that the resonant peak of the cavity experiences a red shift on exposure to the aerosol medium. Besides, when the individual resonant peak is analyzed, we found that the sharpness of the peaks (related to the quality factor, Q), also changes. To analyze the peaks quantitatively, we calculate the Q of the resonant peaks after curve fitting them with an appropriate Lorentz function. The Fig. 6f, 6g, 6h shows the Lorentz fitting on the resonant peaks with 0, 20 and 120 second exposure to the aerosols. \\

We calculate the quality factor of individual resonant peaks obtained from the transmission spectrum of the resonant cavity at two different wavelengths, one absorbing wavelength (i.e. ~1531 nm) where N-methyl aniline absorbs significantly and another non-absorbing wavelength (i.e.~1576 nm) where N-methyl aniline absorption is insignificant. Fig. 6i-j compares the change in quality factor at these wavelengths. A significant reduction in the quality factor at the absorbing wavelength (near 1530 nm) is observed. On the other hand, there is no change for the case of non-absorbing wavelength (near 1576 nm). The relative change in the quality factor provides us the spectral absorption spectrum of the aerosol medium.
The intrinsic loss from the waveguides and noise from other ambient particles in the surrounding medium is characterized. This characterization is used to normalize the quality factor change obtained with the aerosols. \\

\begin{figure}[h!]
	\centering
	\includegraphics[width=1\textwidth]{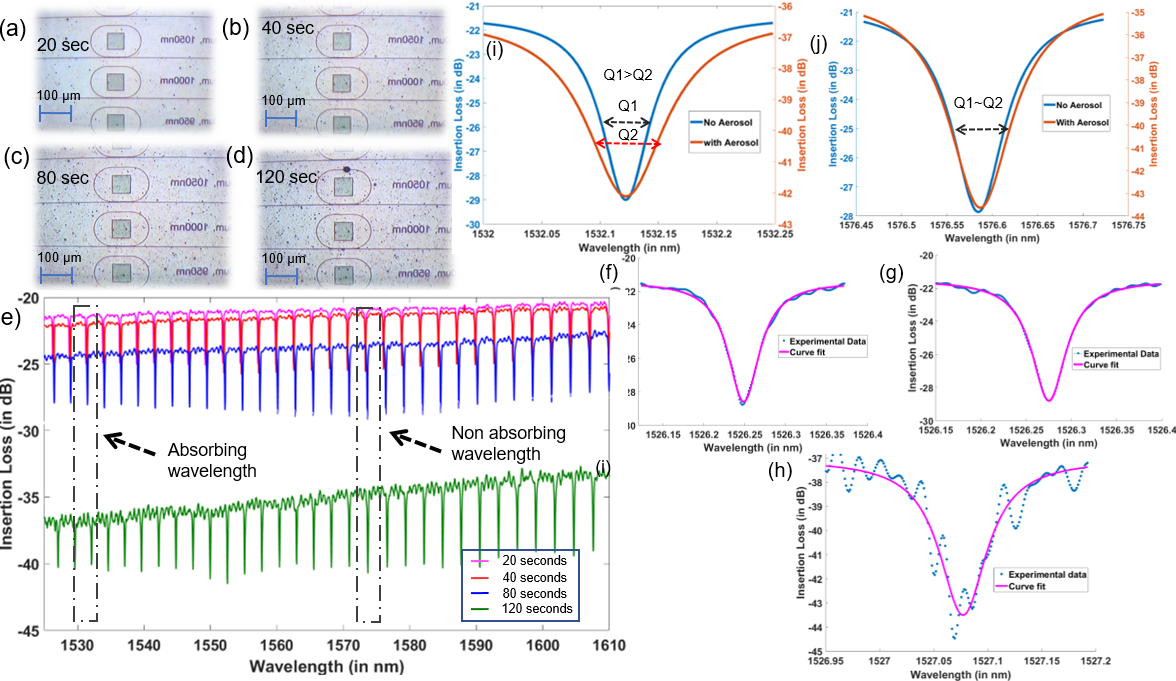}
	\caption{\textit{The microscopic images of the aerosol photonic sensor at different exposure times to the N-methyl aniline aerosol medium (20 sec, 40 sec, 80 sec and 120 seconds). (e) Insertion loss measured with the sensors over the range of the 1530 nm to 1600 nm. (f) Lorentz curve fit plot of the resonant peak with no aerosol exposure (g, h) Lorentz curve fit plot of the resonant peak with aerosol exposure for 20 seconds and 120 seconds respectively (i) Comparison of the quality factor of the resonance peak at the absorbing wavelength. After exposure to the aerosol, the quality factor of the resonance decreases due to light absorption by the interacting particles. (j) Comparison of the quality factor of the resonant peak at a non-absorbing wavelength. The quality factor remains almost the same.}}
\end{figure}

\begin{figure}[h!]
	\centering
	\includegraphics[width=1\textwidth]{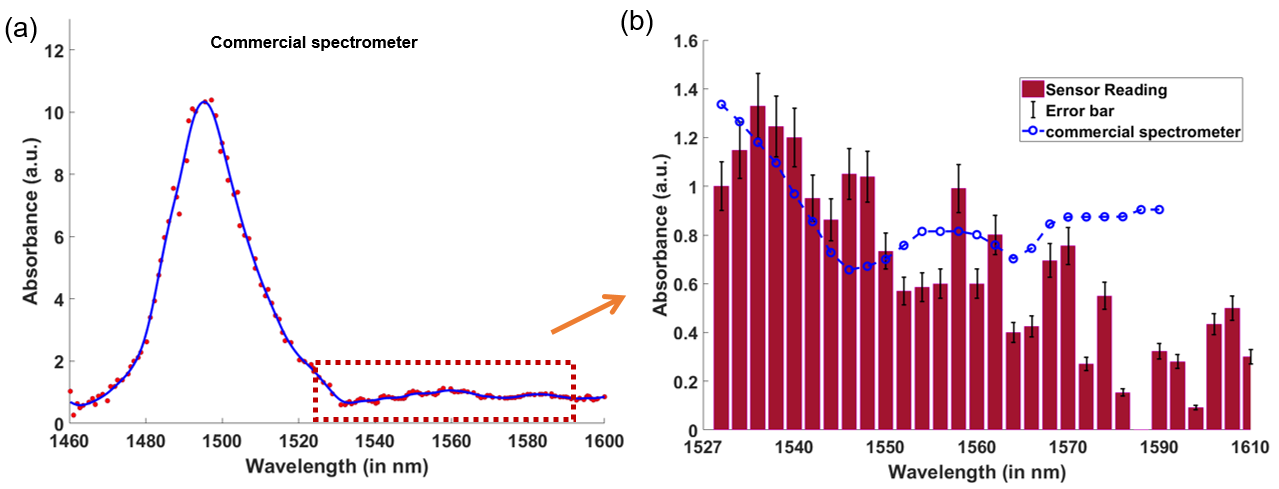}
	\caption{\textit{Absorption spectrum of the N-methyl aniline between 1460 nm to 1600 nm [10]. (b) The absorption spectrum of the N-methyl aniline aerosol obtained from the relative change of quality factor of the resonance peaks in the transmission spectrum of the racetrack resonant cavity. The relative change of the quality factor is obtained by analyzing the Q-factor before and after aerosol exposure to the sensor. We define it as (Q1-Q2)/Q1 where Q1 and Q2 are the quality factors before and after aerosol exposure respectively.}}
\end{figure}

Once we obtain the absorption spectrum from the sensor readings, we compare it with that obtained from a commercial spectrometer \cite{Saito2018}. The commercial spectrometer data is adapted from the experiments performed in the \cite{Saito2018}. N-methyl aniline has an overtone absorption peak between 1460 nm to 1530 nm. Fig. 7 shows the super-imposed spectra from the sensor and commercial spectrometer. A comparison reveals that the spectrum obtained from our sensor matches closely with that of the commercial spectrometer. The error bars on the obtained values were about 10\% of the measured readings. We believe that the spectral resolution of the spectrum can be further improved by using an array of the resonant cavities supporting different wavelengths and providing a number of spectral data points. \\

\section{Conclusion}
The presented work demonstrates the ability to chemically characterize aerosol particles using on-chip photonic cavity enhanced spectroscopy. The cavity is based on the Si3N4 based micro-racetrack resonators. The racetrack cavity improves light coupling efficiency and offers smaller FSR to increase the spectral resolution of the measurements. The photonic cavity increases the effective path length by the order of quality factor of the resonant cavity, making the sensor highly sensitive in sensing and characterizing the aerosol particles. The sensor performs a continuous wavelength sweep detection of the aerosol, from which the absorption spectrum is calculated based on the relative change in quality factor over the selected wavelength range. We illustrate the absorption spectrum retrieval of the N-methyl aniline based aerosols using the method. The results obtained from the proposed method shows a good agreement to the data collected from a commercial spectrometer. We present a sensing modality that is particularly advantageous as a lightweight and smaller footprint device for the accurate characterization of aerosols in remote, on-site, and body wearable sensing applications. Our method overcomes the existing challenge of pre-calibration and pre-knowledge about the aerosol medium in obtaining an absorption spectrum.\\

While the current resonator design and experiments focus on the Near IR range, it is possible to extend the method to mid-IR which is particularly useful to perform chemical spectroscopy. In such cases other mid-IR transparent materials like chalcogenide glasses, etc. can be used to fabricate the device. The system can be miniaturized by using tunable portable on-chip light source and detector. The present method shows a proof-of-concept of cavity enhanced on-chip aerosol spectroscopy.
\bibliographystyle{plain}
\bibliography{paper_bib}

\end{document}